\documentclass[12pt,preprint]{aastex}

\newcommand{\saxj}{SAX~J1808.4--3658}

\newcommand{\xtejb}{XTE~J1807--294}
\newcommand{\xtejc}{XTE~J1814--338}

\newcommand{\nudot}{\dot{\nu}}
\newcommand{\Mdot}{\dot{M}}

\newcommand{\pth}[1]{\left({#1}\right)}
\newcommand{\qpth}[1]{\left[{#1}\right]}

\newcommand{\Sin}[1]{\sin\pth{#1}}

\newcommand{\Tstar}{T_{\star}}
\newcommand{\Porb}{P_{orb}}

\usepackage{subfigure, ulem}
\bibpunct{(}{)}{;}{a}{}{,}

\shorttitle{Timing of XTE~J1807--294}
\shortauthors{Riggio et al.}

\begin{document}

\title{Spin up and phase fluctuations in the timing of the accreting
millisecond pulsar XTE~J1807--294}

\author{A. {Riggio}\altaffilmark{1}, T. {Di Salvo}\altaffilmark{2}, L.
  {Burderi}\altaffilmark{1}, M. T. {Menna}\altaffilmark{3},
  A. {Papitto}\altaffilmark{3}, R. {Iaria}\altaffilmark{2},
  G. {Lavagetto}\altaffilmark{2}}

  \altaffiltext{1}{Dipartimento di Fisica, Universit\`a degli Studi di
  Cagliari, Cittadella Universitaria, S.P. Monserrato - Sestu Km
  0,700 09042, Monserrato (CA), Italy; E-mail: riggio@dsf.unica.it}
  \altaffiltext{2}{Dipartimento di Scienze Fisiche e Astronomiche,
  Universit\`a~di Palermo, Via Archirafi 36, Palermo I-90123, Italy}
  \altaffiltext{3}{Osservatorio Astronomico di Roma, Sede di
  Monteporzio Catone, Via Frascati 33, Rome I-00040, Italy}

\begin{abstract}
  We performed a timing analysis of the 2003 outburst of the accreting
  X-ray millisecond pulsar \xtejb\ observed by RXTE. Using recently
  refined orbital parameters we report for the first time a precise
  estimate of the spin frequency and of the spin frequency
  derivative. The phase delays of the pulse profile show a strong
  erratic behavior superposed to what appears as a global spin-up
  trend.  The erratic behavior of the pulse phases is strongly related
  to rapid variations of the light curve, making it very difficult to
  fit these phase delays with a simple law.  As in previous cases, we
  have therefore analyzed separately the phase delays of the first
  harmonic and of the second harmonic of the spin frequency, finding
  that the phases of the second harmonic are far less affected by the
  erratic behavior. In the hypothesis that the second harmonic pulse
  phase delays are a good tracer of the spin frequency evolution we
  give for the first time a estimation of the spin frequency
  derivative in this source. The source shows a clear spin-up of $\dot
  \nu = 2.5(7) \times 10^{-14}$ Hz sec$^{-1}$ (1 $\sigma$ confidence
  level).  The largest source of uncertainty in the value of the
  spin-up rate is given by the uncertainties on the source position in
  the sky. We discuss this systematics on the spin frequency and its
  derivative.
\end{abstract}

\maketitle

\section{Introduction}
Binary systems in which one of the two stars is a neutron star (NS
hereafter) are among the most powerful X-ray sources of our
Galaxy. The emission of X-rays is due to the matter transferred from
the companion star and accreted onto the NS, and to the release of the
immense gravitational energy during the fall or in the impact with the
NS surface. A sub-category of such systems is called Low Mass X-ray
Binaries (LMXB). LMXBs are characterized by low NS superficial
magnetic fields ($< 10^9~ \mathrm{Gauss}$) and by the low-mass ($< 1
\, M_\odot$) of the companion star. The so-called recycling scenario
\citep[see for a review][]{Bhatta_91} sees in the millisecond radio
pulsars the last evolutionary step of LMXBs, where the torques due to
the accretion of matter and angular momentum, together with the
relatively weak magnetic fields, are able to spin-up the NS to periods
of the order of one millisecond. When the accretion phase terminates
and the companion star stops transferring matter, the NS can switch on
as a millisecond radio pulsar, although no example has been reported
yet.

The recycling scenario received the long awaited confirmation only in
1998 with the discovery of the first millisecond X-ray pulsar in a
transient LMXB; the first LMXB observed to show coherent pulsations at
a frequency of $\sim 400$ Hz was \saxj\ \citep{Wijnands_98}, in which
the NS is orbiting its companion star with a period of $\sim 2.5$ hr
\citep{Chakra_98}.  Why millisecond X-ray pulsars were so elusive is
an argument still debated in literature. A possible reason can be due
to the relatively low magnetic fields of these sources which has
therefore less capability to channel the accreting matter onto the
polar caps, then the chance to see a pulsed emission from a LMXB is
quite low \citep[see e.g.][]{Vaughan_94}, especially at high accretion
rates. However, to date 10 LMXBs have been discovered to be accreting
millisecond pulsars (see \citealp{Wijnands_05}~for a review on the
first 6 discovered, for the last four see \citealp{Kaaret_06,
Krimm_07, Casella_07, Altamirano_07}), and all of them are in
transient systems. They spend most of the time in a quiescent state,
with very low luminosities (of the order of $10^{31} - 10^{32}$
ergs/s) and rarely they go into an X-ray outburst with luminosities in
the range $10^{36} - 10^{37}$ ergs/s. Although the recycling scenario
seems to be confirmed by these discoveries, from timing analysis of
accreting millisecond pulsars we now know that some of these sources
show spin-down while accreting~ \citep{Galloway_02, Papitto_07}. This
means that it is of fundamental importance to study the far from being
understood details of the mechanisms regulating the exchange of
angular momentum between the NS and the accreting matter, and chiefly
the role of the magnetic field in this exchange. The main way to do
this is the study of the pulse phase shifts and their relations with
other physical observable parameters of the NS.

The pulse phase shifts are frequently affected by intrinsic long-term
variations and/or fluctuations (with which we mean an erratic behavior
of the phase delays possibly caused by variations in the instantaneous
accretion torques or movements of the accretion footprints on the NS
surface, see \citealp{DiSalvo_07}\ for a review). Examples of this
complex behavior of the pulse phase shifts in accreting millisecond
pulsars were already reported in literature.  \cite{Burderi_06}, who
analyzed the 2002 outburst of the accreting millisecond pulsar \saxj\
found a jump of 0.2 in the pulse phases of the first harmonic which is
not present in the second harmonic phases, which show a much more
regular behavior. This change is in correspondence of a change in the
slope in the exponential decay of the X-ray light curve (see also
\cite{Hartman_07} for a discussion of the complex phase behavior in
other outbursts of \saxj).  \cite{Papitto_07} found that the second
harmonic of XTE~J1814--338 follows the first harmonic giving
approximately the same spin frequency derivative.  A clear model which
can explain this behavior is still missing, but these observational
evidences seem to suggest that perhaps the second harmonic has a more
fundamental physical meaning. For instance it may be related to the
emission of both the polar caps while the first harmonic may be
dominated by the most intense but less stable polar cap. Another
possible explanation comes from possible shape and/or size variations
of the accretion footprints related to variations of the accretion
rate. \cite{Romanova_03} found a such behavior in their numerical
simulations.

In this paper we report the results of a timing analysis performed on
\xtejb, making use of an improved orbital solution \citep{Riggio_07}.
As in the cases mentioned above, \xtejb\ shows erratic fluctuations of
the phase delays of the first harmonic and a much more regular
behavior of the phase delays derived from the second harmonic. In the
hypothesis that the second harmonic pulse phase delays are a good
tracer of the spin frequency evolution we can derive a spin-up rate of
$2.5(7) \times 10^{-14}$ Hz/s (1 $\sigma$ confidence level).

\section{Observation and Data Analysis}
The millisecond X-ray pulsar \xtejb~was discovered by RXTE on February
21$^{st}$, 2003 \citep{Markwardt_atel_03a}. The source was observed
with the Proportional Counter Array (PCA, $2 - 60$ keV energy range)
and the High Energy X-ray Timing Experiment (HEXTE, $20 - 200$ keV)
on-board of the RXTE satellite \citep{Jahoda_96}, during a long X-ray
outburst which lasted from February 28 to June 22, 2003.  \xtejb~was
also observed with other satellites such as XMM-Newton
\citep{Campana_03, Kirsch_04, Falanga_05}, Chandra
\citep{Markwardt_iauc_03b} and INTEGRAL \citep{Campana_03}. No optical
or radio counterpart have been reported to date.  \cite{Linares_05}
have reported the presence of twin kHz QPOs analyzing RXTE
observations.

Here we analyze all the archival RXTE observations of this source.  In
particular, we use high-time resolution data from the PCA.  We use
data collected in GoodXenon packing mode for the timing analysis,
which permits maximum time and energy resolution (respectively $1 \mu
s$ and 256 energy channels). In order to improve the signal to noise
ratio we select photon events from PCUs top layer and in the energy
range 3-13 keV \citep{Galloway_02}. Indeed, we have verified that this
is the range where we have the highest S/N ratio. In fact, using all
the energy range the pulsations at days 104 and 106, respectively,
after the beginning of the outburst are much less statistically
significant.

Using the \textit{faxbary}\footnote{faxbary is a tool of the HEAsoft
Software Packages. It can be found at:
http://heasarc.nasa.gov/docs/software/lheasoft/} tool (DE-405 solar
system ephemeris) we corrected the photon arrival times for the motion
of the earth-spacecraft system and reported them to barycentric
dynamical times at the Solar System barycenter. We use the source
position reported by \cite{Markwardt_iauc_03b} using a
\textit{Chandra} observation during the same outburst.

To obtain the X-ray light curve during the outburst we used the PCA
data collected in Standard2 mode (256 energy channel and 16s binned
data) and corrected for the background using the faint background
model suitable for the source count rate \citep[see][]{Jahoda_06}. No
energy selection was applied in this case since we are interested in a
good tracer of the bolometric luminosity. We did not apply any
correction for dead time since the maximum count rate was quite low
($< 100$ cts sec$^{-1}$ PCU$^{-1}$, background included); in fact the
mean time between two event is at least two orders of magnitude higher
than the expected dead time ($10 \mu s$) for this count rate
\citep{Jahoda_06}. We selected all the data using both internal GTI
and applying criteria regards pointing offset, South Atlantic Anomaly
passage, electronic contamination and Sun offset\footnote{according to
the prescription given in
http://heasarc.gsfc.nasa.gov/docs/xte/abc/screening.html we adopted as
selection criteria the following: time since SAA greater than 30
minutes, elevation angle with respect the Earth greater than 10
degree, electron contamination lower than 0.1, and pointing offset
lower than 0.25 degree.}.

The resulting light curve is shown in Fig.\ref{fig1} (pentagon
symbols). The flux shows an exponential decay with superimposed six
evident flares. To derive the characteristic time of the exponential
decay we fitted the light curve with an exponential law. In order to
remove the time intervals affected by X-ray flares we excluded from
the fit all the points whose flux was greater than the best fit
exponential model by at least a 15\%. The choice of this threshold is
arbitrary, but a different choice, like 10\% or 20\% include or
exclude very few points. We repeated the exponential fit on the flares
subtracted light curve. In this last fit the $\chi^2/\mathrm{d.o.f.} =
23096/214$ which is awfully high. Such a large $\chi^2$ is due to
deviations of the X-ray light curve from a pure exponential decay (see
e.g.\ all the points after 100 days from the beginning of the
outburst).

Although these deviations may be very small, they can be large if
compared with the statistical error on a single point. However, in
order to obtain a reliable estimate of the parameters of the fit, and
in particular a reliable estimate of the errors, we need to obtain a
reduced $\chi^2$ of the order of unity. Therefore, we have multiplied
the errors on each point by a factor 10. In this way we obtain a
characteristic decay time of $\tau = 17.50(25)$ days.

It should be noted that a constant term must be added to the model to
obtain a good description of the light curve, although background
subtraction was performed. This residual results to be $\sim 10.8(2)$
cts sec$^{-1}$ PCU$^{-1}$ and may be due to a contaminating source in
the PCA field of view. It is unlikely it is due to quiescence emission
since the source was observed in quiescence by XMM Newton and was not
detected \citep{Campana_05}. In both cases this residual flux does not
affect the inferred decaying time of the light curve or any other
results of this paper.

In order to minimize the time delays induced by the orbital motion, we
correct the photon arrival times with the formula \citep[see
e.g.]{Deeter_81}:
\begin{equation}
  \label{eq:time_orb_doppler}
  t_{em} \simeq t_{arr} - A \qpth{\Sin{m(t_{arr}) + \omega} + 
    \frac{\varepsilon}{2} \Sin{2m(t_{arr}) + \omega} - 
    \frac{3\varepsilon}{2}\Sin{\omega}},
\end{equation}
where $t_{em}$ is the photon emission time, $t_{arr}$ is the photon
arrival time, $A$ the projected semi-major axis in light seconds,
$m(t_{arr}) = 2\pi (t_{arr} - \Tstar)/\Porb$ is the mean anomaly,
$\Porb$ the orbital period, $\Tstar$ is the time of ascending node
passage, $\omega$ is the periastron angle and $\varepsilon$ the
eccentricity.  In order to remove completely from the pulse phase
delays any effect due to the orbital motion it is of fundamental
importance to correct the arrival times of the events with very
precise orbital parameters.  To accomplish this task we used the
orbital solution recently published by \cite{Riggio_07}, who, using
the total outburst time available (about 120 days), obtain a solution
that is about two orders of magnitude more precise than previously
reported orbital solutions.

We divided the whole observation in time intervals of length
approximately equal to the orbital period\footnote{This is to minimize
possible residuals due to uncertainties in the orbital parameters,
since we expect these residuals to be periodic at the orbital period
of the system.} and epoch-folded each of these data intervals with
respect to the spin period we reported in Tab.~\ref{table1}. In this
way we are able to significantly detect the X-ray pulsations up to day
106 from the beginning of the outburst, making this as the longest
time span in which timing analysis of an accreting millisecond pulsar
has been performed.

The pulse phase delays are obtained fitting each pulse profile with
two sinusoidal components (with period fixed to 1 and 0.5 of the spin
period, respectively), since the second harmonic was significantly
detected in the folded light curve.  In Fig.~\ref{fig1} and~\ref{fig2}
we show the pulse phase delays of the first harmonic and second
harmonic, respectively. We have plotted only the pulse phase delays
corresponding to the folded light curves for which the statistical
significance for the presence of the X-ray pulsations is $>
3\sigma$. Moreover, we consider the second harmonic significantly
detected (and plotted its phases) only when the ratio between the best
fit amplitude of the second sinusoid and its error was greater than 3
($A/\delta A > 3$).  We propagated on each phase point the errors on
the orbital parameters with the formulas derived by
\citealp{Burderi_07}. We note that the propagated errors in this case,
for which the orbital parameters are known with great precision,
result to be much smaller than the statistical errors derived from the
sinusoidal fit.

As it is evident from Fig.~\ref{fig1} and~\ref{fig2}, the phase delays
of the first harmonic show a noisy behavior with shifts up to 0.3 in
phase.  The noise affecting the phases results strongly
anti-correlated to the source flux, as already noted for another
source of this class \citep{Papitto_07}.  On the other hand, the phase
delays derived from the second harmonic are much more regular, a
behavior that is similar to the one shown by \saxj\
\citep{Burderi_06}. Although a few points (corresponding to rapid
flares in the light curve) appear to be significantly below the
general trend, the phase delays of the second harmonic clearly show a
parabolic decrease, as it is expected in case of a spin-up of the NS.

\section{Timing Results}
Since the phase delays of the second harmonic are much less noisy than
the phases derived from the first harmonic, and assuming that the
pulse phase delays derived from second harmonic are a good tracer of
the spin frequency evolution, we decided to fit the second harmonic in
order to find information on the spin frequency behavior.  To fit the
phase delays we start from the simplest assumption of a constant spin
frequency derivative.  We hence fit the second harmonic phase delays
with the model:
\begin{equation}
  \phi(t) = \phi_0 - \Delta\nu\, (t - T_0) - \frac{\nudot}{2}\, (t-T_0)^2,
\label{eq:model}
\end{equation}
where $T_0$ is the date of the beginning of the observation,
$\Delta\nu$ is a correction to the spin frequency and $\nudot$ the
spin frequency derivative. Using all the data points we obtained a
spin frequency derivative $\nudot = 2.05(28) \times 10^{-14}$ Hz/s
with a quite large $\chi^2/\mathrm{d.o.f.} = 1560.57/142$. From a
visual inspection of the phase residuals with respect to this model
(see Fig.~\ref{fig2}), we can see that the largest contribution to the
$\chi^2$ is given by a group of 3 points at MJD 52713.0 (about 14.5
days from the beginning of the outburst). These points (indicated with
triangles in Fig.~\ref{fig2}) correspond to the largest flare visible
in the light curve and to a strong decrease of the phases of the first
harmonic as well (cf.\ Fig.~\ref{fig1}). We therefore believe that
this is a phase shift induced by the rapid change of the X-ray flux
similar to the phase shifts observed in the first harmonic. If we
remove from our data set these three points (case B) and re-perform
the fit, we obtain a frequency derivative $\nudot = 2.26(15) \times
10^{-14}$ Hz/s, perfectly compatible with the value previously found,
demonstrating that the three points we have eliminated do not affect
the spin frequency derivative obtained by the fit. In this case of
course the statistical quality of the fit increases, giving a
$\chi^2/\mathrm{d.o.f.} = 452.4/139$.  However, this $\chi^2$ is still
unacceptable; again the post-fit residuals indicate that the major
contribution to the $\chi^2$ is given by all the points in
correspondence of the X-ray flares.  We have therefore decided to
remove all the points (indicated with circles in Fig.~\ref{fig2}) that
fall in time intervals for which the flux results to be larger by 15\%
with respect to the exponential best fit function derived above.  In
this way total of 21 points were excluded from the fit (case C). With
this last data set we obtained a value of spin frequency derivative
$\nudot = 2.46(15) \times 10^{-14}$ Hz/s (again compatible with the
results obtained with the complete data set) and a
$\chi^2/\mathrm{d.o.f.} = 257.6/121$.  In this case, a value of $\nu =
190.623507018(6)$ Hz for spin frequency at the beginning of the
outburst was obtained.

We have also tried to fit this (reduced) data set with a spin-up model
which takes into account the decrease of the X-ray flux (supposed to
trace the mass accretion rate) during the outburst (see
\citealp{Burderi_06} for a more detailed discussion). In principle
this correction should be important for this source, given the
particularly long duration of the outburst (about 120 days). Fitting
the phase delays of the second harmonic with eq.~1 of
\cite{Burderi_06}, in which we adopted an exponential decay time of
the X-ray flux of 17.50(25) days, as derived from the X-ray light
curve, we obtain a significant improvement of the fit with a
$\chi^2/\mathrm{d.o.f.} = 225.5/121$ (a $\Delta \chi^2 = 32$ for the
same number of degrees of freedom). In this case, we obtain a spin
frequency derivative at the beginning of the outburst of $\dot \nu_0 =
1.25(7)\times 10^{-13}$ Hz/s, corresponding to a mass accretion rate
at the beginning of the outburst of $\dot M_0 = 4.03(23) \times
10^{-10}$ M$_\odot$/yr, and a best fit spin frequency of $\nu_0 =
190.623506939(7)$ Hz. In Fig.\ref{fig3} we report, among the last
reduced data set, both parabolic and exponential best fit models and
the residuals of the exponential model (bottom panel).

Unfortunately, the results above are affected by large systematic
uncertainties, given by the large uncertainty on the source
coordinates (which is about $0''.4 (1\sigma \; \textrm{confidence
level})$ from a Chandra observation\footnote{The Chandra observation
of \xtejb\ in outburst was performed with the instrument HRC-S. As
reported in http://asc.harvard.edu/cal/ASPECT/celmon/, the confidence
levels are given at 68\% ($0''.4$), 90\% ($0''.6$) and 99\%
($0''.8$).}), that we are now going to discuss in detail.

The uncertainties in the phase delays caused by the uncertainties on
the estimates of the source position in the sky, will produce a
sinusoidal oscillation at the Earth orbital period. For observation
times shorter than one year, as it is the case for most transient
accreting millisecond pulsars, this can cause systematic errors on the
determination of the NS spin period and its derivative, since a series
expansion of a sinusoid contains a linear and a quadratic term. In the
case of \xtejb, due to the low positional precision
\citep{Markwardt_atel_03b} and the long time span in which the
pulsation is visible (up to 106 days from the beginning of the
outburst), we obtain, from the expression given by \cite{Burderi_07},
the following systematic uncertainties in the spin frequency and the
spin frequency derivative, respectively: $\sigma_{\nu\, {\rm pos}}
\sim 4.1 \times 10^{-8}$ Hz and $\sigma_{\dot \nu\, {\rm pos}} \sim
0.8 \times 10^{-14}$ Hz/s. Since this error is of the same order of
magnitude of our best fit estimate of $\nudot$, we need to evaluate
these effects in a more careful manner.

Let us consider the expression of the phase delays induced by the
Earth motion for a small displacement, $\delta \lambda$ and $\delta
\beta$, in the position of the source in ecliptic coordinates,
$\lambda$ and $\beta$ \citep[see e.g.][]{Lyne_90}:
\begin{equation}
\label{eq:posdel}
\Delta \phi_{\rm pos}(t) = \nu_0\, y \,\left[\,\sin (M_0 + \epsilon)
\cos \beta\, \delta \lambda - \cos (M_0 + \epsilon) \sin \beta
\,\delta \beta \,\right]
\end{equation}
where $y = r_{\rm E}/c$ is the distance of the Earth with respect to
the Solar system barycenter in light seconds, and $M_0 = 2 \pi (T_0 -
T_\gamma)/P_\oplus - \lambda$, where $T_0$ is the begin of the
observation, $P_\oplus$ the Earth orbital period, $T_\gamma$ the time
of passage through the Vernal point, and $\epsilon = 2 \pi (t -
T_0)/P_\oplus$.  As already done by \cite{Burderi_07},
Eq.~\ref{eq:posdel} can be rewritten as:
\begin{equation}
\label{eq:posdel2}
\Delta \phi_{\rm pos} = \nu_0\, y \,\sigma_\gamma \, u \,\Sin{M_0 + 
\epsilon - \theta^*}
\end{equation}
where $\sigma_\gamma$ is the positional error circle, $\theta^* =
\arctan (\tan \beta \; \delta \beta / \delta \lambda)$, and $u =
[(\cos \beta\; \delta \lambda)^2 + (\sin \beta\; \delta \beta)^2\;
]^{1/2} / \sigma_\gamma$. We can safely pose $u = 1$ as an upper
limit.

In order to take into account the effects of an incorrect source
position, we fitted the reduced data set (case C) with a model which
takes also into account the modulation caused by the incorrect source
coordinates and given by Eq.~\ref{eq:posdel2}:
\begin{equation}
  \phi(t) = \phi_0 - \Delta\nu\, (t - T_0) - \frac{\nudot}{2}\, 
(t - T_0)^2 + \Delta \phi_{\rm pos}(t)
\label{eq:model2}
\end{equation}
We have repeated the fit changing $\sigma_\gamma$ and $\theta^*$ in
such a manner to cover the Chandra error box up to a 90 \% confidence
level, that is sky region within an angular distance of $0.''6$ from
the reported source position.  The obtained values of the spin
frequency and its derivative for each possible position of the source
within the Chandra error box are shown in Fig.~\ref{fig4}. The values
of $\dot \nu$, at $1\sigma$ confidence level, range in the interval
$(1.8 - 3.2) \times 10^{-14}$ Hz s$^{-1}$, while the best fit value of
the frequency derivative for the source nominal position is $2.46(15)
\times 10^{-14}$ Hz s$^{-1}$.  It is evident that the effect of the
poor source position knowledge is much larger than the statistical
error on the parabolic fit. Still the spin-up behavior of the source
remains significant even considering the large uncertainties caused by
the position uncertainties.

A similar discussion must be done for the spin frequency. The best fit
value for the nominal position is $\nu =190.623507018(4)$ Hz, while
the variation of the linear term in the fit at different positions of
the source inside the Chandra error box are in the range $\Delta\nu =
\pm 4 \times 10^{-8}$ Hz, one order of magnitude greater than the
single fit statistical error. Finally, the reduced $\chi^2$ for these
fits varies in the range (2.1 - 2.4).

Summarizing, using the pulse phase delays derived from the second
harmonic, we inferred the spin frequency derivative in \xtejb. In the
hypothesis of constant spin frequency derivative we obtain a value of
$\nudot = 2.46(15) \times 10^{-14}$ Hz s$^{-1}$. In the hypothesis of
an exponential decay of the accretion rate we obtained a value of the
spin frequency derivative at the beginning of the outburst of $\dot
\nu_0 = 1.25(7)\times 10^{-13}$ Hz s$^{-1}$. These results do not
include the systematic errors induced by the poorly constrained source
position. Taking into account the errors on the source position we
obtained, for the constant and exponential decay models, respectively,
the values of $2.5(7) \times 10^{-14}$ Hz s$^{-1}$ and $ 1.25(33)
\times 10^{-13}$ Hz s$^{-1}$.

\section{Discussion and conclusion}
We have analyzed a long RXTE observation of the accreting millisecond
pulsar \xtejb\ and reported the results of an accurate timing analysis
performed on a time span of about $120$ days, the longest outburst of
an accreting millisecond pulsar for which a timing analysis has been
performed to date.  We find that the phase delays derived from the
first harmonic show an erratic behavior around a global parabolic
spin-up trend.  This behavior is similar to that previously shown by
two accreting millisecond pulsar, \saxj\ \citep{Burderi_06} and
\xtejc\ \citep{Papitto_07}. In the case of the 2002 outburst of \saxj,
the phase delays of the first harmonic show a shift by about 0.2 in
phase at day 14 from the beginning of the outburst, when the X-ray
flux abruptly changed the slope of the exponential decay.  On the
other hand, the phase delays of the second harmonic in \saxj\ showed
no sign of the phase shift of the first harmonic, and could be fitted
by a spin-up in the first part of the outburst plus a barely
significant spin-down at the end of the outburst.  In the case of
\xtejc, the fluctuations in the phase delays were visible both in the
first harmonic and in the second harmonic, superposed to a global
parabolic spin-down trend. \cite{Papitto_07} have shown that the
post-fit phase residuals were strongly anti-correlated to variations
of the X-ray light curve. These fluctuations were interpreted as due
to movements of the accretion footprints (or accretion column) induced
by variations of the X-ray flux.

\begin{figure}
  \plotone{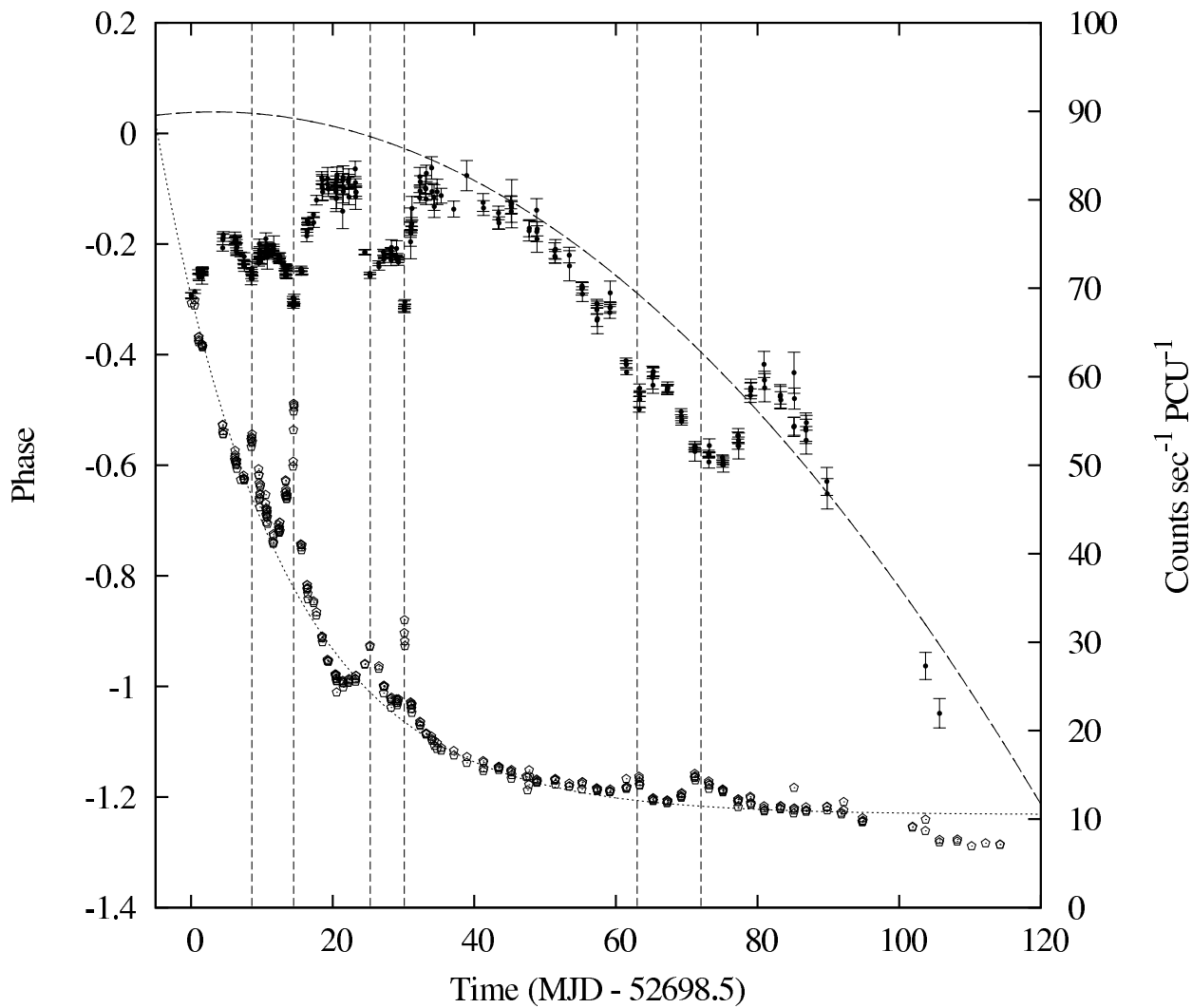}
  \caption{Light curve of \xtejb\ during the 2003 outburst (pentagon)
    and phase delays of the first harmonic as a function of time
    (small dot).  The dashed vertical lines indicate the times of six
    clearly visible flares of the X-ray flux superimposed to a global
    exponential decay. The dotted curve represent the exponential fit
    of the light curve, obtained after having previously excluded from
    the data the six flares. The dashed curve represent the parabolic
    best fit obtained fitting the second harmonic phase delays and
    considering the nominal source position. Strong fluctuations of
    the phase delays are apparent and are strongly anti-correlated to
    the flares present in the X-ray light curve.\label{fig1}}
\end{figure}

\begin{figure}
  \plotone{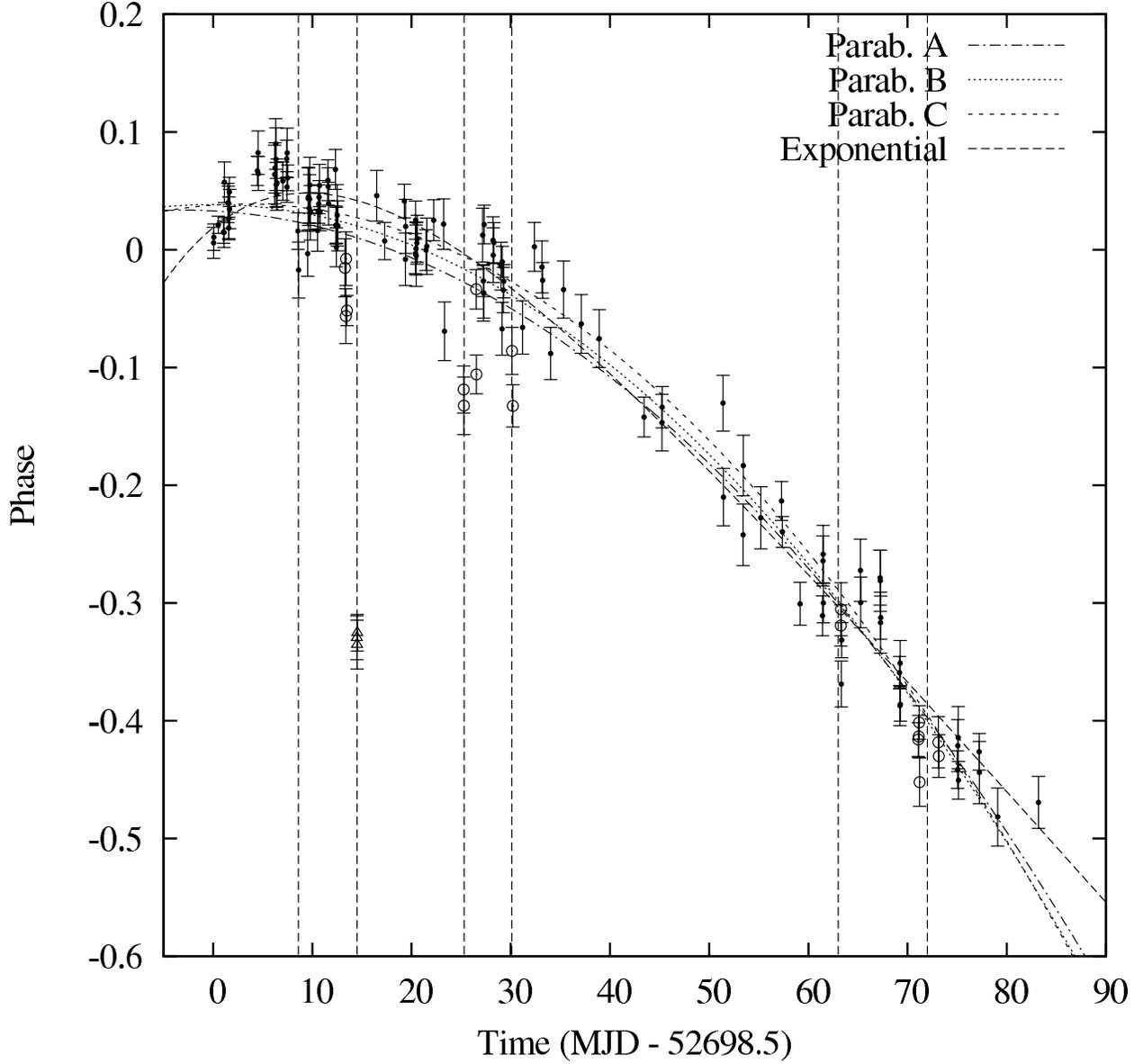}
  \caption{ Plot of \xtejb\ second harmonic pulse phase delays.  The
    four curves represent the parabolic best fit for the nominal
    source position, respectively, using all the data points (case A),
    excluding the three point at MJD 52713.0 (case B, where the points
    excluded are identified by triangles), and excluding all the data
    points for which the flux exceeds the best-fit exponential decay
    for more than 15\% (Case C, where the points excluded are
    identified by circles), and the best fit obtained using an
    exponentially decreasing mass accretion rate (see text). The
    exponential fit was performed on the data sub-set corresponding to
    case C.\label{fig2}}
\end{figure}

\begin{figure}
  \plotone{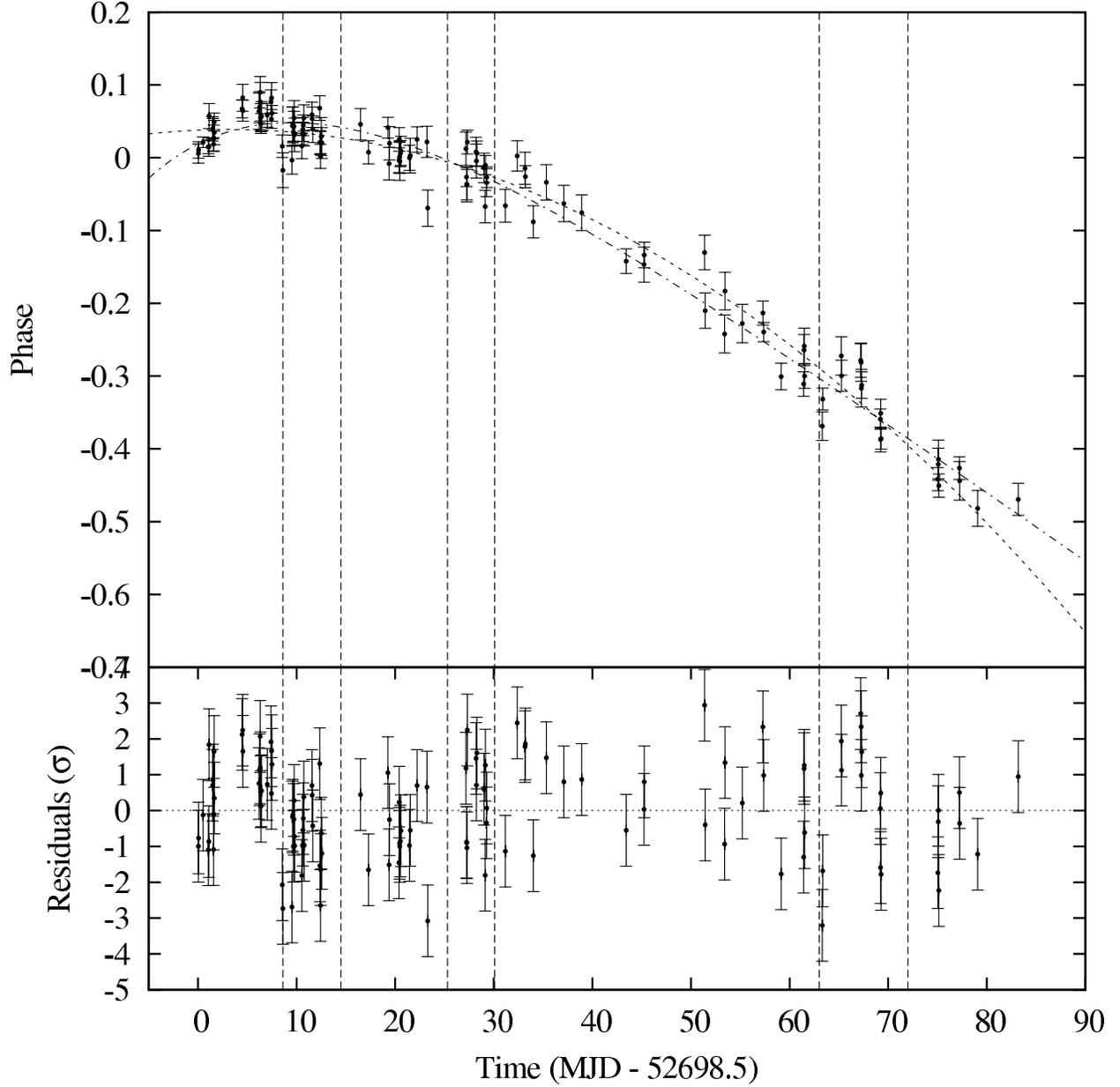}
  \caption{second harmonic pulse phase delays together with the
    parabolic and exponential best fit (top panel), and residuals in
    units of $\sigma$ with respect to the exponential best fit model
    (bottom panel) considering only the sub-set of case C (see
    Fig.\ref{fig2}).\label{fig3}}
\end{figure}

\begin{figure}
%  \epsscale{.80}
  \plottwo{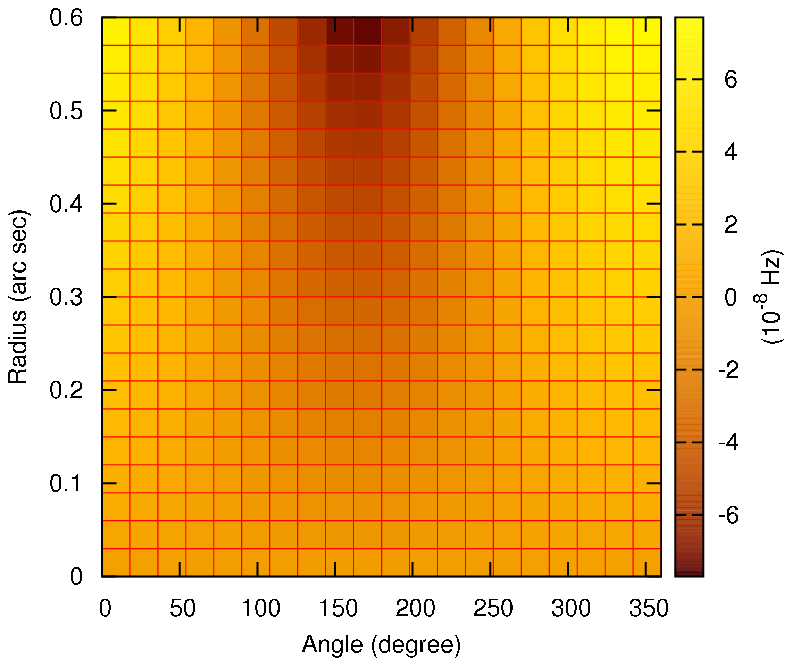}{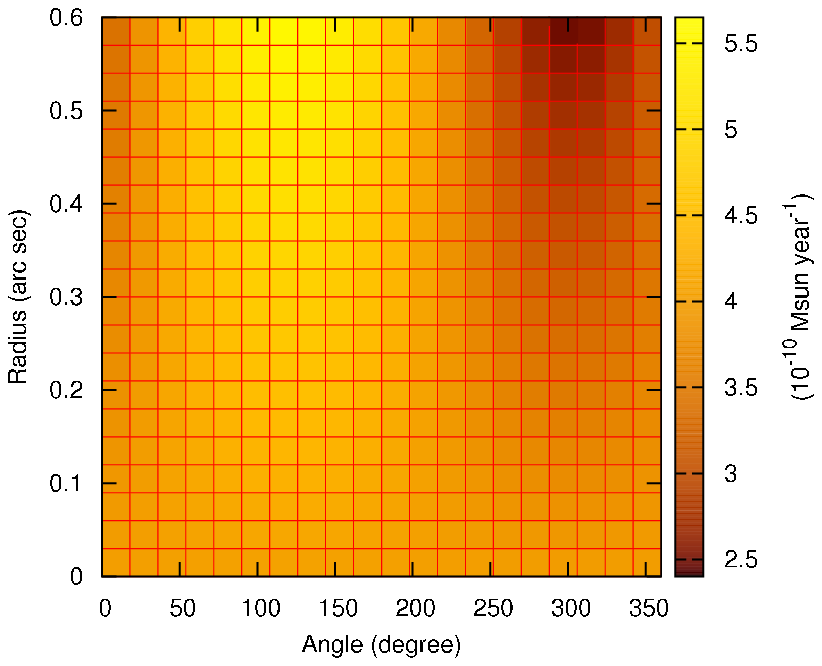}
  \caption{Diagrams of the best fit values of $\Delta \nu$ (left
    panel) and $\Mdot$ (right panel) obtained fitting the first
    harmonic pulse phase delays with the expression \ref{eq:model2},
    as function of the parameters $\sigma_\gamma$ and $\theta^*$ (see
    text).\label{fig4}}
\end{figure}

In the case of \xtejb, the fluctuations in the phase delays affect
mostly the first harmonic, which shows a trend that is very difficult
to reproduce with a simple model.  As in the case of \xtejc, the
post-fit phase residuals are clearly anti-correlated with variations
observed in the X-ray light curve; from Figure~\ref{fig1} we see that
the phases decrease when the X-ray flux shows rapid increases. It is
important to note that the anti-correlation visible between the
post-fit phase delays and the X-ray flux is independent of the
spin-down or spin-up behavior of the source, since it is observed in
\xtejc, which shows spin-down, and in \xtejb, which shows spin-up.
The correlation between the phase delays and the X-ray flux affects
the second harmonic only marginally.  Indeed, there are a few points
in the phase delays of the second harmonic that are significantly
below the global trend observed in the phase delays, and all of them
correspond to flares in the X-ray light curve. Excluding these points
marginally affects the values we obtain for the spin frequency and its
derivative, but gives a significant improvement of the $\chi^2$ of the
fit.

We find that the phase delays of the harmonic are fitted by a
parabolic spin-up model. We have also showed that the quality of the
fit is much improved if we use a more physical model in which the
spin-up rate decreases exponentially with time following the decrease
of the X-ray flux (and hence of the inferred mass accretion rate).  In
fact, if the spin-up of the source is related to the mass accretion
rate, then it should not be constant with time, but, in first
approximation, should decrease proportionally with the mass accretion
rate onto the NS. For instance, assuming that accretion of matter and
angular momentum occurs at the corotation radius $R_{co}$, the
relation between the spin frequency derivative and the mass accretion
rate is, from the angular momentum conservation law, $\nudot = \Mdot
\, \sqrt{G M R_{co}}/ 2 \pi I$, where G is the gravitational constant,
M the NS mass and I is the NS moment of inertia; this gives a lower
limit on the mass accretion rate since the specific angular momentum
at the corotation radius is the maximum that can be transferred to the
NS. In the case of \xtejb, the duration of the outburst is
particularly long (about 120 days), and the effect of the global
decrease of the mass accretion rate during the outburst should be
particularly relevant for this source. Indeed in this case the fit we
obtain using an exponentially decreasing spin-up rate is significantly
better than using a constant spin-up rate.

From the fit of the phase delays of the second harmonic of \xtejb\
with the model discussed above we find a mass accretion rate at the
beginning of the outburst of $4(1) \times 10^{-10}$ M$_\odot$
yr$^{-1}$.\footnote{For this estimation we adopted the value of
$I=10^{45}$ g cm$^2$, M = 1.4 M$_\odot$ and NS radius $R_{NS} = 10^6$
cm.} We can compare this mass accretion rate with the X-ray flux of
the source at the beginning of the outburst that was $2 \times
10^{-9}$ ergs cm$^{-2}$ s$^{-1}$ \citep{Falanga_05} from which we
derive an X-ray luminosity of $4.7 \times 10^{36}$ ergs s$^{-1}$ and a
distance to the source of 4.4(6) kpc. Clearly this is only a crude
estimation of the distance on the basis of our timing results and
future independent estimates are needed to confirm or disprove our
hypothesis.

\acknowledgements This work was supported by the Ministero della
Istruzione, della Universit\`a e della Ricerca (MIUR), national
program PRIN2005 2005024090$\_$004.

% Stile bibliografico e bibliografia
\bibliography{ms}

\begin{deluxetable}{lc}
  \tablecaption{Orbital and Spin Parameters for XTE J1807-294.}
  \label{table1}
  \tablehead{ \colhead{Parameter} & \colhead{Value}}
  \startdata
  RA (J2000) & \(18^h\, 06^m\, 59.8^s \)\tablenotemark{a}  \\
  Dec (J2000) & \(-29^o\, 24'\, 30"\)\tablenotemark{a} \\
  Orbital period, $P_{orb}$ (s) & 2404.41665(40)\tablenotemark{b} \\
  Projected semi-major axis, $a_x \sin i$ (lt-ms) & 4.819(4)\tablenotemark{b} \\
  Ascending node passage, T$^\star$~\tablenotemark{a} (MJD) & 52720.675603(6)\tablenotemark{b} \\
  Eccentricity, e  & $<$ 0.0036\tablenotemark{b} \\
  Reference epoch, T$_0$~\tablenotemark{c} (MJD) & 52698.5 \\
  \textit{Parabolic fit results} &  \\
  Spin frequency, $\nu_0$ (Hz) & 190.62350702(4) \\
  Spin frequency derivative, $\nudot$ (Hz s$^{-1}$) & $ 2.5(7) \times 10^{-14}$ \\
  \textit{Exponential fit results} & \\
  Spin frequency, $\nu_0$ (Hz) & 190.62350694(5) \\
  Spin frequency derivative, $\nudot$ (Hz s$^{-1}$) & $ 1.25(33) \times 10^{-13}$
  \enddata

  \tablecomments{Errors on orbital parameters are intended to be at
                 $1\sigma$ confidence level (c. l.), upper limits are
                 given at 95\% c.l. Best fit spin parameters are
                 derived in both hypothesis of a constant spin-up and
                 flux dependent spin-up, and the uncertainties include
                 systematics due to the uncertainties in the source
                 position (see text).}
  
  \tablenotetext{a}{\cite{Markwardt_iauc_03b}.}
  \tablenotetext{b}{\cite{Riggio_07}.}  
  \tablenotetext{c}{This is the Epoch at which are referred the
  reported values of $\nu$ and $\nudot$.}
\end{deluxetable}

\end{document}